\documentclass[twocolumn,pra,longbibliography,superscriptaddress]{revtex4-1}

\DeclareMathAlphabet{\mymathbb}{U}{BOONDOX-ds}{m}{n}

\usepackage{amsthm}
\usepackage{graphicx}
\usepackage{amsmath}

\usepackage{amsfonts}
\usepackage{amssymb}

\usepackage{amsbsy}
\usepackage{siunitx}
\newtheorem{theorem}{Theorem}

\newcommand{\iu}{{i\mkern1mu}}
\newcommand{\s}{\mathsf{S}}
\newcommand{\Ll}{\mathsf{Y}}

\newcommand{\ANU}{Centre for Quantum Computation and Communication Technology, Department of Quantum Science,
Research School of Physics and Engineering, Australian National University, Canberra ACT 2601, Australia.\looseness=-1}
\newcommand{\NTU}{Nanyang Quantum Hub, School of Physical and Mathematical Sciences, Nanyang Technological University, Singapore 639673.\looseness=-1}
\newcommand{\NTUc}{Complexity Institute, Nanyang Technological University, Singapore 639673.\looseness=-1}

\newcommand{\TJU}{College of Precision Instrument and Opto-Electronics Engineering, Key Laboratory of
Opto-Electronics Information Technology, Ministry of Education, Tianjin University, Tianjin 300072, China.\looseness=-1}

\newcommand{\NUS}{Centre for Quantum Technologies, National University of Singapore,
3 Science Drive 2, Singapore, Republic of Singapore.\looseness=-1}

\begin{document}

\title{Decoupling Cross-Quadrature Correlations using Passive Operations}

\author{Syed M. Assad}
\email{cqtsma@gmail.com}
\affiliation{\NTU}
\affiliation{\ANU}

\author{Mile Gu}
\email{gumile@ntu.edu.sg}
\affiliation{\NTU}
\affiliation{\NUS}
\affiliation{\NTUc}

\author{Xiaoying Li}
\affiliation{\TJU}

\author{Ping Koy Lam}
\affiliation{\ANU}
\affiliation{\NTU}

\date{August 13, 2020}

\begin{abstract}
  Quadrature correlations between subsystems of a Gaussian quantum
  state are fully characterised by its covariance matrix. For example,
  the covariance matrix determines the amount of entanglement or
  decoherence of the state. Here, we establish when it is possible to
  remove correlations between conjugate quadratures using only passive
  operations. Such correlations are usually undesired and arise due to
  experimental cross-quadrature contamination. Using the
  Autonne--Takagi factorisation, we present necessary and sufficient
  conditions to determine when such removal is possible. Our proof is
  constructive, and whenever it is possible we obtain an explicit
  expression for the required passive operation.
\end{abstract}

\maketitle

\section{Introduction} The decomposition of Gaussian quantum systems
has proven to be a fruitful subject of research. For instance, the
textbook examples of Williamson~\cite{williamson,Simon1994} and
Braunstein~\cite{braunstein} tell us that any Gaussian state can be
decomposed through beamsplitters, phase shifters and single-mode
squeezers into uncorrelated thermal states. This is useful for
designing quantum gates~\cite{shiozawa}. More generally, instead of
demanding the complete diagonalisation of the state, it can also be
transformed into another that has specific kinds of
correlations. Early examples of this are the Simon and Duan {\em et
  al.}  standard forms~\cite{simon2000,duan}: using local squeezing
and phase shifts to bring an entangled state into some standard form
of correlations. This turned out to be important in advancing our
understanding of Gaussian entanglement.

All the transformations above require the use of active operations and
bring the state to a form that does not have any cross-quadrature
correlations. Active operations are those that require an external
source of energy, for example, squeezing, while passive operations are
those that do not~\cite{Weedbrook2012}.  Active operations are usually
more difficult to implement in a real device compared to passive
operations which can be implemented almost free of errors using
beamsplitters and phase shifts~\cite{Reck1994}. When restricted to
only passive operations, a generic Gaussian state cannot be
diagonalised; it can only be brought to standard forms that remain
correlated. There exist conditions with which one can check whether a
Gaussian state can be diagonalised by a passive
operation~\cite{Simon1994,Arvind1995}. These conditions are always
satisfied when the Gaussian states are
pure~\cite{braunstein,Arvind1995}.

Here, instead of requiring the state to be fully diagonalised, we
report a necessary and sufficient condition under which the
correlations between conjugate quadrature variables can be entirely
removed using passive operations only. This is stated in the following
theorem.
\begin{theorem}\label{thm:main}
  Let
  $\mathbf{a}=\left[a_1,\ldots,a_n,a_1^\dagger,\ldots,a_n^\dagger
  \right]$ be a vector collecting the annihilation and creation
  operators of $n$ modes. Let
  \begin{equation*}
    \s_{jk} = \frac{1}{2} \mathrm{Tr}\left[\rho \left(\mathbf{a}_j \mathbf{a}_k^\dagger +\mathbf{a}_k^\dagger \mathbf{a}_j\right) \right]= \begin{bmatrix}
  \mathsf{X} & \mathsf{Y} \\
  {\mathsf{Y}}^* & {\mathsf{X}}^*
\end{bmatrix}_{jk}
  \end{equation*}
  be the complex covariances of an $n$-mode Gaussian state $\rho$
  having zero mean $\mathrm{Tr}\left[\rho\, \mathbf{a} \right]=0 $.
  Then $\s$ can be brought into a cross-quadrature decorrelated form
  using passive operations if and only if there exist an
  Autonne--Takagi factorisation of $\mathsf{Y}$:
  $\mathsf{Y} = \mathsf{Z}^\dagger\Ll_0 \mathsf{Z}^*$, and a diagonal
  matrix $\mathsf{R}$ with entries in $\{1,i\}$ such that
  $\mathsf{R}^\dagger\mathsf{Z}\mathsf{\mathsf{X}}\mathsf{Z}^\dagger\mathsf{R}$ is
  real. Furthermore, the required passive operation is given by
  $\mathsf{Z}$ up to swapping of quadratures determined by
  $\mathsf{R}$.
\end{theorem}

The crux of the theorem is the diagonalisation of $\mathsf{Y}$, which
is given to us by the Autonne--Takagi factorisation~\cite{autonne,
  takagi}.
\begin{theorem}[Autonne--Takagi factorisation]\label{thm:at}
  Let $\mathsf{Y}$ be a complex symmetric matrix. Then there
  exists a unitary matrix $\mathsf{Z}$ such that $\mathsf{Y}
  =\mathsf{Z}^\dagger \Ll_0 \mathsf{Z}^*$, with $\Ll_0$ real,
  non-negative and diagonal. 
\end{theorem}

The diagonal entries of $\mathsf{Y}_0$ are the singular values of $\mathsf{Y}$
in any desired order. The uniqueness property of $\mathsf{Z}$ is
stated in Appendix~\ref{AT-decomposition}. Essentially, the physical
situation of interest is a correlated state with unwanted correlations
between some of the conjugate quadratures and we are concerned with
the conditions under which these unwanted correlations can be removed
using only passive operations. We mean ``conjugate quadratures'' in a
more general sense---\textit{any} quadrature pairs, $q_j$ and $p_k$
with $j$ not necessarily equal to $k$ and where
$[q_j,p_k]=\iu\delta_{jk}$. In other words, theorem 1 identifies
those states that are composed of $q$correlations and
$p$ correlations plus passive operations. As a corollary, it also
identifies states which cannot be constructed by passive operations on
initially uncorrelated, squeezed or otherwise, single modes. The proof
of the theorem is constructive in that the required passive operation
is obtained whenever it exists. It turns out to be, up to local
rotations, just $\mathsf{Z}$ given by the Autonne--Takagi's factorisation,
which is very convenient.

We note that Autonne--Takagi's factorisation makes its appearance in
multimode quantum optics~\cite{cariolaro, arzani} that resembles the
approach we have taken here, but there is one important
difference---we consider the factorisation of quantum states rather
than the decomposition of unitaries for determining supermodes as is
the case in multimodal theories.

\section{Proof of theorem 1}
In what follows, we prove Theorem~\ref{thm:main}. We work
with the complex covariance matrix which can be obtained
from the quadrature covariance matrix by the change of variables~\cite{Schumaker1985}
\begin{align}
  \label{eq:qtoa}
  a_j= \frac{q_j + \iu p_j}{\sqrt{2}}\; \text{ and }\;    a_j^\dagger =\frac{q_j -\iu p_j}{\sqrt2}\,.
\end{align}
The reason for working in such a basis is twofold. First, the
conjugate quadratures have vanishing correlations if and only if both
matrices $\mathsf{X}$ and $\mathsf{Y}$ are real. Second, passive
operations take the simple form
\begin{align*}
  \begin{bmatrix}
    \mathsf{E} & 0\\
    0 & \mathsf{E}^*
  \end{bmatrix}
\end{align*}
with $\mathsf E$ unitary due to the symplectic conditions. A direct
calculation shows that the covariance matrix transforms as
$\mathsf{E}:(\mathsf{X},\mathsf{Y}) \mapsto
(\mathsf{E}\mathsf{X}\mathsf{E}^\dagger,\mathsf{E}\mathsf{Y}\mathsf{E}^\intercal)$
under passive operations, whence it follows that the problem of
decoupling conjugate variables is reduced to finding a unitary matrix
$\mathsf{E}$ such that $\mathsf{E}\mathsf{X}\mathsf{E}^\dagger$ and $\mathsf{E}\mathsf{Y}\mathsf{E}^\intercal$ are simultaneously real.
We can now proceed to prove the main result.

\begin{proof}[Proof: Forward direction]
  Suppose $\s$ is the covariance matrix of a state $\rho$ with cross-quadrature
  correlations which can be removed by a passive operation
  $\mathsf{Q}$. In other words, after applying $\mathsf{Q}$, the
  cross-quadrature correlations 
  $\left\{ q_j, p_k\right\}=0$, where to
  simplify notations, we use
$\left\{ q_j, p_k\right\} $ to mean $\frac12 \text{Tr} \left[ \rho( q_j p_k+p_k
    q_j)\right]$.
    In the complex representation,
  denoting the transformed matrix as
  $\mathsf{X}_1=\mathsf{Q}\mathsf{X}\mathsf{Q}^\dagger$ and
  $\mathsf{Y}_2=\mathsf{Q}\mathsf{Y}\mathsf{Q}^\intercal$,
  the transformed covariance matrix has entries
  \begin{align*}
   [\mathsf{X}_1]_{jk}=\left\{ a_j ,a_k^\dagger\right\} &=   \frac{ \left\{ q_j, q_k
                                       \right\}}{2} +
                                       \frac{\left\{ p_j, p_k \right\}}{2} \\
    [\mathsf{Y}_2]_{jk}= \left\{ a_j, a_k \right\}&=  \frac{  \left\{ q_j, q_k
                                        \right\} }{2}-
                                        \frac{\left\{ p_j, p_k \right\}}{2}
  \end{align*}
  which are real. Since $\mathsf{Y}_2$ is a real symmetric
  matrix, it has a spectral decomposition
  $\mathsf{Y}_2 =\mathsf{R}_1^\intercal \Ll_1
  \mathsf{R}_1$~\cite{horn2012}, where $\mathsf{R}_1$ is a real
  orthogonal matrix and $\Ll_1$ is a real (but not necessarily
  positive) diagonal matrix the entries of which are the eigenvalues of
  $\mathsf{Y}_2$. To obtain the Autonne--Takagi decomposition, consider
  a passive unitary (but not necessarily real) transformation
  $\mathsf{R}:(a_j,a_j^\dagger ) \mapsto ( ia_j,-i
    a_j^\dagger)$ on $\mathsf{Y}_1$ for every $j\in J$ where $J$
  is the set containing all indices $j$ for which $[\Ll_1]_{jj}$ is
  negative.  This corresponds to a rotation of the quadratures
  $\mathsf{R}:(q_j,p_j)\mapsto (p_j,-q_j)$ for $j\in J$. In matrix
  form, $\mathsf{R}$ is diagonal with entries
  \begin{align*}
    [\mathsf{R}]_{jk}=\begin{cases}
      1&\text{for } j=k \notin J \,,\\
      i&\text{for } j=k \in J \,\\
      0&\text{for } j \neq k\,.
    \end{cases}
  \end{align*}
  Applying this to $\Ll_1$ brings it to a non-negative diagonal matrix
  $\Ll_0= \mathsf{R} \Ll_1 \mathsf{R}^\intercal$
  since
  \begin{align*}
    \mathsf{R}: \left\{ a_j, a_j \right\} \mapsto \begin{cases}
      -\left\{a_j ,a_j \right\}& \text{ for } \left\{  a_j ,a_j
      \right\}<0\,,\\
      \left\{a_j ,a_j \right\}& \text{ for } \left\{  a_j ,a_j
      \right\}\geq0\,.
      \end{cases}
  \end{align*}
 Putting everything
  together, we arrive at the Autonne--Takagi decomposition of $\mathsf{Y}$ as
  \begin{align*}
    \mathsf{Y}= \underbrace{\mathsf{Q}^\dagger \mathsf{R}_1^\intercal \mathsf{R}^\dagger}_{\mathsf{Z}^\dagger} \Ll_0 \underbrace{\mathsf{R}^*
     \mathsf{R}_1 \mathsf{Q}^*}_{\mathsf{Z}^*}\,.
  \end{align*}
  Then $\mathsf{X}$ transforms as
  \begin{align*}
    \mathsf{Z}\mathsf{X}\mathsf{Z}^\dagger &= \mathsf{R} \mathsf{R}_1^* \mathsf{Q}\mathsf{X} \mathsf{Q}^\dagger
                                             \mathsf{R}_1^\intercal \mathsf{R}^\dagger\\
&=    \mathsf{R} \underbrace{\mathsf{R}_1^* \mathsf{X}_1 \mathsf{R}_1^\intercal}_{\mathsf{X}_0} \mathsf{R}^\dagger\,,
  \end{align*}
where $\mathsf{X}_0$ is a real (symmetric) matrix since both
$\mathsf{X}_1$ and $\mathsf{R}_1$ are real. This implies
$\mathsf{R}^\dagger\mathsf{Z}\mathsf{X}\mathsf{Z}^\dagger\mathsf{R}$
is real which completes the proof.
\end{proof}

\begin{proof}[Proof: Reverse direction]
  Let $\mathsf{Z}$ be the unitary matrix in the Autonne--Takagi
  factorisation of $\mathsf{Y}$:
  $\mathsf{Y}=\mathsf{Z}^\dagger\mathsf{Y}_0\mathsf{Z}^*$ and
  $\mathsf{R}$ be a diagonal matrix with entries in $\{1,i\}$ such that
  $\mathsf{R}^\dagger\mathsf{Z}\mathsf{X}\mathsf{Z}^\dagger\mathsf{R}$
  is real. The passive transformation
  $\mathsf{R}^\dagger\mathsf{Z}$ results in
  $\mathsf{R}^\dagger\mathsf{Z}:(\mathsf{X},\mathsf{Y})\mapsto(\mathsf{R}^\dagger\mathsf{Z}\mathsf{X}\mathsf{Z}^\dagger\mathsf{R},
  \mathsf{R}^\dagger\mathsf{Z}\mathsf{Y}\mathsf{Z}^\intercal\mathsf{R}^*)$. The
  first term is real by assumption. The second term
  \begin{equation*}
     \mathsf{R}^\dagger\mathsf{Z}\mathsf{Y}\mathsf{Z}^\intercal\mathsf{R}^*
     = \mathsf{R}^\dagger\mathsf{Z}\mathsf{Z}^\dagger \mathsf{Y}_0 \mathsf{Z}^*\mathsf{Z}^\intercal\mathsf{R}^*= \mathsf{R}^\dagger\mathsf{Y}_0\mathsf{R}^*
  \end{equation*}
  is also real since $\mathsf{Y}_0$ is a real diagonal matrix. When
  $\mathsf{X}$ and $\mathsf{Y}$ are simultaneously real, it follows
  from direct substitution that the quadrature covariance matrix has no
  cross-quadrature correlations.
\end{proof}  

\begin{figure}[!t]
\centering
\includegraphics[width=0.9\columnwidth]{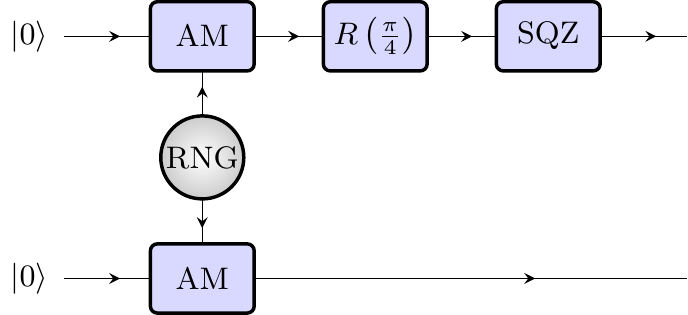}
\caption{The output state with quadrature
  covariance matrix given by~(\ref{cccstate}) has cross-quadrature correlations that
  cannot be removed by passive operations. AM: Amplitude
  modulator. RNG: Gaussian random number generator with variance
  $1/2$. $R(\frac{\pi}{4})$: $\pi/4$ phase shifter. SQZ: \SI{3}{\dB} squeezer.}
\label{fig:1}
\end{figure}

What does this mean? It means that we have a way of testing if the
correlations between conjugate variables can be removed---diagonalise
$\mathsf{Y}$ to obtain the matrix $\mathsf{Z}$ using the
Autonne--Takagi factorisation and subsequently compute
$\mathsf{Z}\mathsf{X}\mathsf{Z}^\dagger$. If $\mathsf{Y}$ is a
full-rank matrix with non-degenerate eigenvalues and
$\mathsf{Z}\mathsf{X}\mathsf{Z}^\dagger$ cannot be transformed to a
real matrix by a diagonal matrix $\mathsf{R}$, then the correlations
cannot be decoupled. This is certainly the case if
$\mathsf{Z}\mathsf{X}\mathsf{Z}^\dagger$ has any entries that are
neither real nor purely imaginary. On the other hand, if all the
entries of $\mathsf{Z}\mathsf{X}\mathsf{Z}^\dagger$ are real, then
$\mathsf{Z}$ is the passive operation that we are after. If some
entries are purely imaginary then in addition to $\mathsf{Z}$,
additional local rotations $\mathsf{R}$ are required. If $\mathsf{Y}$
is singular or has degenerate eigenvalues, then we have some freedom
in choosing $\mathsf{Z}$ to make the entries of
$\mathsf{R}^\dagger\mathsf{Z}\mathsf{X}\mathsf{Z}^\dagger\mathsf{R}$
real.

When $\mathsf{S}$ corresponds to a {\em pure state}, the matrix
$\mathsf{Z}$ gives the passive operation required to create it from a
product of independent squeezed states. However, if $\mathsf{S}$ is
mixed, our result implies that it is sometimes impossible to create by
passive operations on {\em any} independent states, or even on states
possessing only $q$ correlations and $p$ correlations. One example is
the state with quadrature covariance matrix
  \begin{align}
    \label{cccstate}
\mathcal{S}=\frac{1}{2} \begin{bmatrix}
  3 & 0.5 & 1 & 0\\
  0.5 & 0.75 & 0.5 & 0\\
  1 & 0.5 & 2 & 0\\
  0 & 0 & 0 &1
\end{bmatrix}
\end{align}
which can be created by the scheme in Fig~\ref{fig:1}. The squeezing
operation ``locks in'' the cross-quadrature correlations and makes it
impossible to be removed using passive operations only.

\section{Two-mode example}
We illustrate our result by working through an example. Consider a
two-mode Gaussian state having the following quadrature
covariance matrix
\begin{align*}
\mathcal{S}=\begin{bmatrix}
  m & 0 & c & 0\\
  0 & m & 0 & -c\\
  c & 0 & n & s\\
  0 & -c & s & n
\end{bmatrix}
\end{align*}
with all $m$, $n$, $c$ and $s$ positive. We want to determine if this
state can be brought into a cross-quadrature decorrelated form. The basis
transformation~(\ref{eq:qtoa}) represented by the unitary matrix
\begin{align*}
\mathcal{L}= \frac{1}{\sqrt{2}} \begin{bmatrix}
  1 & \iu & 0 & 0\\
  0 & 0 & 1 & \iu\\
  1 & -\iu & 0 & 0\\
  0 & 0 & 1 & -\iu
\end{bmatrix}
\end{align*}
transforms the quadrature covariance matrix into the complex
covariance matrix
\begin{align*}
  \mathsf{S}=\mathcal{L} \mathcal{S} \mathcal{L}^\dagger
  =\begin{bmatrix}
  m & 0 & 0 & c\\
  0 & n & c & \iu s\\
  0 & c & m & 0\\
  c & -\iu s & 0 & n
\end{bmatrix}\,,
\end{align*}
which identifies $\mathsf{X}$ and $\mathsf{Y}$ as
\begin{align*}
  \mathsf{X} =\begin{bmatrix}
  m & 0 \\
  0 & n 
\end{bmatrix}\;\text{ and }
      \mathsf{Y} =\begin{bmatrix}
  0 & c \\
  c & \iu s 
\end{bmatrix}\,.
\end{align*}
The Autonne--Takagi factorisation of
$\mathsf{Y}=\mathsf{Z}^\dagger\Ll_0\mathsf{Z}^*$ is given
by
\begin{align*}
\mathsf{Z} &=e^{i \pi/4} \begin{bmatrix}
  -i\sqrt{t} & \sqrt{1-t} \\
  \sqrt{1-t} & -i\sqrt{t} 
\end{bmatrix}
\end{align*}
and
\begin{align*}
\Ll_0 &=\frac{1}{2}\begin{bmatrix}
    \sqrt{4c^2+s^2}-s & 0\\
 0&  \sqrt{4c^2+s^2}+s
\end{bmatrix}
\end{align*}
with $t = (1+s/\sqrt{4c^2 +s^2})/2$. This results in
\begin{align*}
  \mathsf{Z}\mathsf{X} \mathsf{Z}^\dagger =
  \begin{bmatrix}n(1-t)+m t&-\iu \sqrt{t(1-t)}(m-n)\\
    \iu \sqrt{t(1-t)}(m-n)& n t+m(1- t)
    \end{bmatrix}
\end{align*}
which has entries that are all real or purely imaginary, and is
transformed to a real matrix by 
\begin{align*}
  \mathsf{R} =\begin{bmatrix}
  1 & 0 \\
  0 & \iu 
\end{bmatrix}
\end{align*}
so that finally we have
\begin{align*}
 \mathsf{R}^\dagger \mathsf{Z}\mathsf{X} \mathsf{Z}^\dagger \mathsf{R}=
  \begin{bmatrix}n(1-t)+m t&\sqrt{t(1-t)}(m-n)\\
    \sqrt{t(1-t)}(m-n)& n t+m(1- t)
    \end{bmatrix}\,.
\end{align*}
This means that the state $\mathsf{S}$ can be brought to a cross-quadrature
decorrelated form and the passive operation that does this is
$\mathsf{R}^\dagger \mathsf{Z}$. This can be factorised as
\begin{align*}
 \mathsf{R}^\dagger \mathsf{Z}=
  \begin{bmatrix}e^{\iu \pi/4}&0\\
    0&e^{-\iu 3\pi/4}
  \end{bmatrix}
        \begin{bmatrix}\sqrt{t}&\sqrt{1-t}\\
    -\sqrt{1-t}&\sqrt{t}
  \end{bmatrix}
        \begin{bmatrix}e^{-\iu \pi/2}&0\\
    0&1
    \end{bmatrix}
  \end{align*}
which is realised by a beamsplitter of transmissivity $t$ and three phase shifts: $\pi/4$
and $-3\pi/4$ at the outputs and $-\pi/2$ at the input port.

The expert reader might have recognised that the state $\mathsf{S}$
can in fact be cross-quadrature decorrelated through the simpler
transformation
\begin{align*}
 \mathsf{R}^\dagger \mathsf{Z}=
  \begin{bmatrix}e^{\iu \pi/4}&0\\
    0&e^{-\iu \pi/4}
  \end{bmatrix}
\end{align*}
requiring just two phase shifts. This shows that when it
is possible to decorrelate the conjugate quadratures the procedure we
presented is not the only way to do so. The condition that
$\mathsf{Y}$ be diagonalised can be relaxed---all we need to decouple
$q$ and $p$ is for $\mathsf{Y}$ to be transformed into a real matrix
after applications of the passive operation---this real matrix need not
be diagonal or non-negative. In terms of implementations, this would mean
that the required operation might be simpler, for instance we can do
away with the beamsplitter in the example considered.

\section{Discussions} An immediate application theorem 1 is to
the calculation of the ``squeezing of formation''~\cite{idel}. This
quantity measures how much squeezing is required to create a given
state and indicates the degree of nonclassicality of the state.
Squeezing of formation is invariant under passive operations because
these transformations do not require any squeezing. This means that
the result of this paper can be used to simplify complicated states to
a form in which the squeezing of formation can be directly
calculated. For example, a brute force computation of the squeezing of
formation for a two-mode Gaussian state involves an optimisation over
six free parameters. However, by first transforming the state to a
quadrature-decorrelated form, if it is possible, this computation
reduces to a simple one parameter optimisation
problem~\cite{tserkis2020}.

There is also an interesting connection with the generation of cluster
states. A cluster state has multiple quantum modes with correlations
between each mode~\cite{menicucci, raussendorf1, raussendorf2}. Many
of these can be shown to possess correlations only between the $q$'s
and between the $p$'s, such as the two-dimensional square
cluster. However, in real devices for generating cluster states there
are imperfections which give rise to correlations between $q$ and
$p$. This implies that our result might be useful for identifying if
an ideal cluster state can be recovered using only passive operations.

What can be said about a state with cross-quadrature correlations
which cannot be removed by passive operations? While most theoretical
work on Gaussian quantum information consider cross-quadrature
decorrelated states, almost every state realised experimentally would
have some cross-quadrature correlations that cannot be decoupled using
only passive operations. However, if we are also allowed to add
correlated noise in the form of random Gaussian quadrature displacements,
then any state can be cross-quadrature decorrelated.  One obvious
question is then the following: what is the least amount of noise required to
achieve such decorrelation?

\begin{acknowledgments}
  We acknowledge H. Jeng for preparing an earlier version of the
  paper. We thank B. Shajilal, T. Michel and S. Tserkis for useful
  discussions. This work is supported by the Australian Research
  Council under the Centre of Excellence for Quantum Computation and
  Communication Technology (Grants No. CE110001027, No. CE170100012,
  and No. FL150100019), the National Research Foundation
  (NRF). Singapore, under its NRFF Fellow programme (Award
  No. NRF-NRFF2016-02), the Singapore Ministry of Education Tier 1
  Grant No. MOE2017-T1-002-043, Grant No FQXi-RFP-1809 from the
  Foundational Questions Institute and Fetzer Franklin Fund (a
  donor-advised fund of Silicon Valley Community Foundation).
\end{acknowledgments}

\appendix
\section{Uniqueness of Autonne--Takagi decomposition}
\label{AT-decomposition}
For completeness, this Appendix recalls the uniqueness properties of
the Autonne--Takagi decomposition. See for example the textbook
by~\citet{horn2012} for proofs.

Let $\mathsf{Y}$ be an $n\times n$ complex symmetric matrix of rank
$r$. Let $\lambda_1,\ldots,\lambda_d$ be the distinct positive
singular values of $\mathsf{Y}$, in any given order with respective
multiplicities $n_1,\ldots,n_d$. Let
$\mathsf{Y}_0=\lambda_1 \mymathbb{1}_{n_1}\oplus \ldots \oplus
\lambda_d \mymathbb{1}_{n_d}\oplus \mymathbb{0}_{n-r}$; the zero block
is missing if $\mathsf{Y}$ is nonsingular. Let $\mathsf{U}$ and
$\mathsf{V}$ be unitary. Then the Autonne--Takagi decomposition of
$\mathsf{Y}$:
$\mathsf{Y}=\mathsf{U}\mathsf{Y_0}\mathsf{U}^\intercal =
\mathsf{V}\mathsf{Y_0}\mathsf{V}^\intercal $ if and only if
$\mathsf{V}= \mathsf{U Q}$, with
$\mathsf{Q}=\mathsf{Q}_1\oplus \ldots \oplus \mathsf{Q}_d \oplus \mathsf{W}$
where each $\mathsf{Q}_j$ is an $n_j \times n_j$ real orthogonal
matrix and $\mathsf{W}$ is an $(n-r)\times(n-r)$ unitary matrix. If
the singular values of $\mathsf{Y}$ are distinct (that is, if
$d\geq n-1$), then $\mathsf{V}=\mathsf{UD}$, in which
$\mathsf{D}=\text{diag}(d_1,\ldots,d_n)$ with $d_j=\pm 1$ for each
$j=1,\ldots,n-1$. The last entry  $d_n=e^{i \theta}$ if $\mathsf{Y}$
is singular ($d=n-1$), otherwise $d_n=\pm 1$ if $\mathsf{Y}$ is nonsingular
($d=n$).

\bibliography{bibfile.bib}
\end{document}